\documentclass{article}
\usepackage[english]{babel}
\usepackage[letterpaper,top=2cm,bottom=2cm,left=3cm,right=3cm,marginparwidth=1.75cm]{geometry}

\DeclareUnicodeCharacter{0301}{*************************************}
\DeclareUnicodeCharacter{0300}{*************************************}

\usepackage{amsmath}
\usepackage{graphicx}
\usepackage{authblk}
\usepackage[colorlinks=true, allcolors=blue]{hyperref}
\usepackage{subfig}
\usepackage{multirow}
\usepackage{caption}
\usepackage{array}
\usepackage{booktabs}

\author[1,3]{Rania Abdelghani}
\author[1]{Pierre-Yves Oudeyer}
\author[2]{Edith Law}
\author[3]{Catherine de Vulpillières}
\author[1]{Hélène Sauzéon}
\affil[1]{Flowers team, Inria Bordeaux, France}
\affil[2]{HCI lab, Waterloo University, ON, Canada}
\affil[3]{EvidenceB, Paris, France}

\title{Conversational agents for fostering curiosity-driven learning in children}

\begin{document}
\maketitle

\begin{abstract}
Curiosity is an important factor that favors independent and individualized learning in children. Research suggests that it is also a competence that can be fostered by training specific metacognitive skills and information-searching behaviors. In this light, we develop a conversational agent that helps children generate curiosity-driven questions, and encourages their use to lead autonomous explorations and gain new knowledge. The study was conducted with 51 primary school students who interacted with either a neutral agent or an “incentive” agent that helped curiosity-driven questioning by offering specific semantic cues. 
Results showed a significant increase in the number and the quality of the questions generated with the "incentive" agent. This interaction also resulted in longer explorations and stronger learning progress. 

Together, our results suggest that the more our agent is able to train children's curiosity-related metacognitive skills, the better they can maintain their information-searching behaviors and the more new knowledge they are likely to acquire.
\end{abstract}


\section{Introduction}
Teaching strategies that rely on giving direct instructions to children have long been used in our classrooms and proven to be rather effective with learning standardized pedagogical content. Yet it has been shown that children learn and memorize the information better when they look for it themselves \cite{Jirout0}. Such curiosity-driven practices can also offer the individualized and independent learning environments that today's education systems need. Cranton~\cite{Cranton0} suggests that epistemic curiosity (i.e the desire for knowledge that motivates learning new material) can be fostered in classrooms by providing activating events and time for puzzlement and critical self-reflections. Jirout~\cite{Jirout0} also suggests that promoting curiosity comes from promoting comfort with uncertainties and encouragement to think about what one does not know, to ask questions and to explore one’s environment. The importance of curiosity has gathered much attention in research, as it leads to a better learning experience and to the raising of the child’s cognitive awareness of his own knowledge. 

Indeed, research, such as in \cite{Bethier0}, shows that the brain is made to memorize by asking questions and not simply by receiving information through reading or listening to it. These studies show that students who ask questions and make hypotheses perform much better than those to whom the information was simply passed. On another hand, Berlyne~\cite{Berlyne0} explains that being aware of one's own missing knowledge is the crucial factor that incites individuals to seek information. Taken together, these findings lead us to suggest that training the curiosity-driven information-seeking behaviors should start with training the metacognitive skill of identifying one's knowledge limits and needs. 

\begin{figure}[hp]
\centering
\captionsetup{justification=centering,margin=1cm}
\fbox{
\includegraphics[width=1\linewidth]{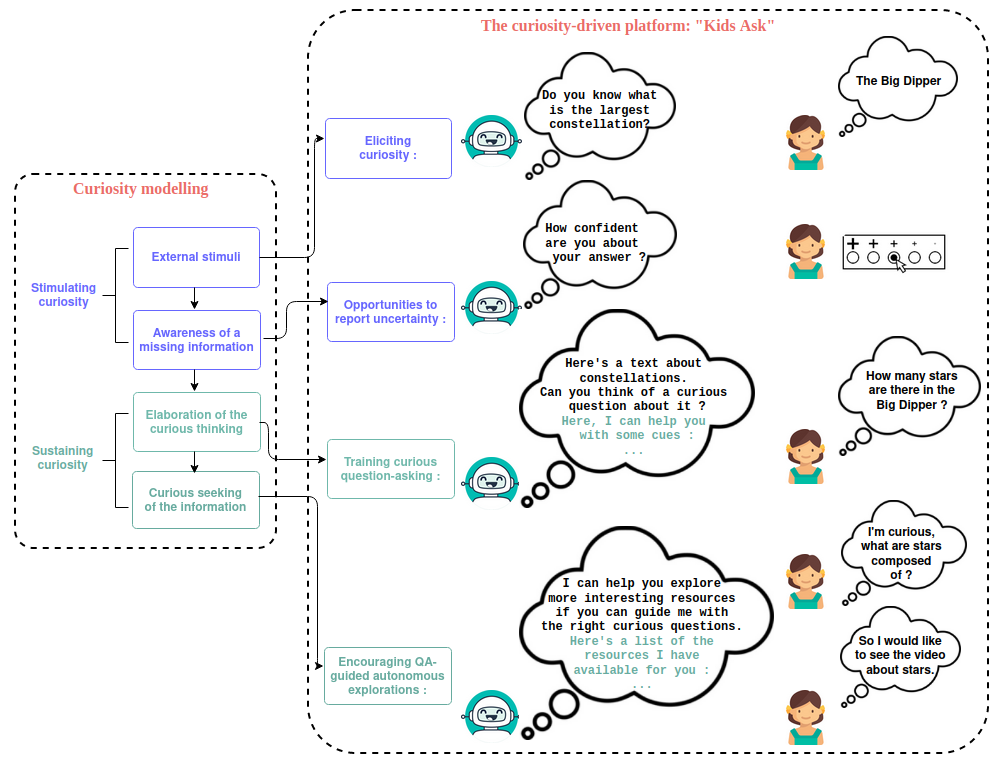}}
\caption{\textbf{Illustration of the study's general idea and the platform's main features}}
  \label{fig:study-ig}
\end{figure}

In this work, we take inspiration from curiosity models and frameworks in order to design a a novel educational platform called "Kids Ask" involving a conversational agent that trains curiosity-based question-asking skills and encourages their use to lead autonomous explorations of their learning environments. We also investigate the role of these curious behaviors in helping children acquire new knowledge (see Figure\ref{fig:study-ig}). Beyond exploring the use of conversational agents only to train curiosity-driven question-asking skills, our study goes one step further and investigates also how such agents can help children use this skill to move forward with their learning. It contributes both to the ongoing research on novel learning approaches in general and to our understanding of how to design effective interactive platforms for promoting curiosity-driven learning.

\section{Related work}
\subsection{Epistemic curiosity, meta-cognition and information-seeking behaviors}

The main event that triggers individuals' information-seeking behaviors is them realizing that they are missing this information. Loewenstein explains this idea in his "information gap" theory \cite{Loewen0} where he says that the activation of epistemic curiosity depends directly on the individual's awareness of his learning processes and how close he feels to closing a knowledge gap: “seeking out new knowledge is more motivated by the aversiveness of not possessing the information than it is by the anticipation of pleasure from obtaining it”. Oudeyer and Kaplan, in \cite{oudeyer0}, propose to complete this cycle with their Learning Progress Theory, suggesting that, to continue seeking information, individuals need to work on activities that provide them with a feeling of an optimal learning progress. Ultimately, Murayama~\cite{Murayama0} presents curiosity as an intrinsically-motivated, goal-oriented cycle that results in information-searching behaviors and that is sustained by a continuous need to optimize the reward felt when acquiring new information. In \cite{Murayama2}, authors argue that it is also possible to feel a strong motivational urge for information, even in the absence of the expected pleasant experience that will emerge from acquiring it. They show that, together, the information’s inherent reward and the cognitive desire for it can predict whether or not the information-seeking behaviors will be happening. Collectively, these theories suggest that training children’s curiosity-driven learning should focus on training the cognitive mechanisms involved in its initiation and sustenance, such as the awareness of a knowledge-gap and how to use the appropriate strategies to resolve one.

\subsection{Divergent questions and exploration during learning}
\label{section: div-ques}
In~\cite{Graesser0}, authors investigate what motivates people to generate questions. Most importantly to our study, they highlight an epistemic curiosity-related factor: a deficit in knowledge and a need to compensate for it. Besides pinpointing the importance of question-asking in curiosity-driven processes, this finding also suggests the need to differentiate between question types: curiosity is more associated with questions that lead to learning new information than with those that are only used to scan explicit facts. 

For this, Bereither~\cite{Bereiter0} proposed two types of questions children can generate in a reading-comprehension task: text-based and knowledge-based. The first type are questions which answers can be directly derived from the text, while the second type reflects the child’s cognitive desire to know or make sense of something. Although text-based questions are easier to think of, knowledge-based questions remain more interesting for learning as they involve higher-order thinking and more cognitive processes like the integration of newly-acquired information into one’s knowledge base. Similarly, Gallagher and al.~\cite{Ascher0} propose to classify questions as divergent- and convergent-thinking. The first being surface-level questions that require children to explain or compare ideas while the second involve divergent thinking processes and require prediction, making hypotheses or judgements. Authors in \cite{Alaimi0} investigated the effect of curiosity on students' questioning abilities and found that the more curious a student is by trait, the more divergent-thinking questions he will ask. 

On another note, in studying the role of question-asking during exploration and learning, researchers in \cite{Vale0} propose that asking a question and pursuing its answer is a fuel for inquiry-based and independent learning. Indeed, according to \cite{Kintsch0}, finding the answer to a question requires the individual not only look for the information, but also to make inferences and use already-acquired knowledge in new situations. For example, in \cite{Bulgren0}, authors developed a routine where participants learn by asking and exploring: they were first asked to formulate a 'critical' question related to the learning material. They were then required to explore the available resources and formulate subsequent questions. Finally, they were asked to use these questions and explored resources to formulate a concise answer to their 'critical' question. Results showed significant differences in learning progress between the students that were trained with this routine and those who had a traditional lecture session. These findings suggest that the explicit thinking of a question can positively affect students' answer-searching behaviors and learning performances.

Despite the important role of generating high-order questions during learning, this behavior is still almost absent from today’s classrooms. In \cite{Graesser0}, the authors explain this by children’s tendency to overestimate their knowledge: children don’t ask questions because they don’t know that they don’t know. On another hand, Humphries and al. suggest in \cite{Humphries0} that children don’t ask higher-order questions because they don’t know how to formulate them. In fact, their results showed that children tended to ask text-based and convergent-thinking questions because they are easier to construct. However, the results also suggest that the ability of asking divergent-thinking questions can be mentored through demonstrations and invitations to deeper thinking. Another brake keeping children from asking questions \cite{Post0} is their negative perception of asking questions and the fear of classmates' judgement.

\subsection{Technology applications for priming curiosity-driven behaviors during learning}
Research has shown that the social environments have a paramount influence on children’s exploratory behaviors and motivation to learn \cite{Engel0}. In this light, several studies have investigated the possibility to use social artificial agents to prime curiosity-driven behaviors in children. In Ceha et al. \cite{ceha0}, children interacted with a social robot that either expressed curiosity, curiosity plus rationale, or no curiosity during a rock classification game. Their results showed that children were able to recognize the robot’s curiosity and to ‘catch’ its curiosity-driven emotions and behaviors. But in terms of intrinsic motivation, authors were not able to assess a significant difference between the conditions with respect to enjoyment, competence, effort and relatedness. In Gordon's work \cite{Goren0}, children interacted with an autonomous agent that exhibited curiosity-driven behaviors. Results showed that the interaction selectively increased children’s aspects of curiosity behavior, but led to no significant differences in learning progress, leaving the question of `whether curiosity can lead to learning gains` still unanswered. In another study \cite{Goren1}, Gordon also presents a question-asking based exploration game. The results showed that the precision of the questions that children generated was a significant predictor for their curiosity, as reported by their teachers. In addition, in Sher and al.'s work in \cite{Sher0}, participants used a web-based application that favoured autonomous exploratory behaviors. Their results showed that enabling time to explore resulted in longer exploration and better learning of the self-explored knowledge, suggesting that exploration improves independent learning. On another note, Clément and al. \cite{Clement0} used the learning progress theory of curiosity to adapt learning tasks to each child and showed that students were more motivated to learn when the lessons were scheduled by curiosity-driven algorithms rather than with the predefined learning sequences. Finally, Alaimi and al. \cite{Alaimi0} developed a pedagogical agent to train children to generate either convergent or divergent questions, using specific semantic cues. These cues consisted of giving the answers to either divergent or convergent questions and asking participants to formulate the questions that lead to them. Their results showed that the divergent-thinking oriented support helped children increase their number of higher-level questions. In addition, and although they did not perceive a significant effect of the participants' perception of curiosity, they found a positive effect of this latter on the children's ability to generate divergent questions. As promising as their results are, the agent here does not move forward with its training to show children how to use this question-asking skill during their learning. They also do not inform us about the impact of this training on the children's domain-knowledge learning achievements.

\subsection{Current study}
Previous work motivates our study as it shows that technology-mediated solutions for enhancing curiosity-driven learning are a promising path. However, the question remains unanswered as to how could children use curiosity-based behaviors as a strategy to optimize their learning. Thereby, our study aims to design an attractive technology-mediated curiosity training and analyze its effect on children's learning strategies and progress. With this mindset, we implement a conversational agent that trains divergent-thinking questioning through fostering knowledge-gap awareness. It also encourages using this skill to lead intrinsically-motivated explorations that help gain new information. The agent finally proposes evaluations to help investigate the effect of the training on the domain-knowledge learning progress.

\section{Study design}
\subsection{Kids Ask}
"Kids Ask" is a novel web-based educational platform that involves an interaction between a child and a conversational agent. The platform is designed to teach children how to generate curiosity-based questions and use them in their learning in order to acquire new knowledge.

To meet this aim, "kids Ask" offers the three following workspaces:
\begin{itemize}
    \item \textbf{The question-asking training space (The QA training space)}: In this space, the child has different texts relating to a theme of his choice. For each text, he interacts with the agent that will try to help him think of divergent questions by giving him specific cues. These cues are either a questioning word (example: 'what difference'), or a questioning word combined with an answer to a possible curiosity-driven question about the text (example: 'what difference' + 'The vaccine prevents the disease, the medication treats it'). The child is then asked by the agent to use the cue(s) to find the corresponding question (for the example above, the question could be 'What is the difference between vaccines and medication?'). The agent chooses to give one or both cues depending on the condition the child is assigned to: control or experimental (see section \ref{section: conditions} for more details). 
    \item \textbf{The exploration space}: In this space, the child has a library of animated educational videos; only 3 of these are initially accessible. Every time the child watches a video, the agent reappears and asks him to generate a divergent question about the video he just seen, without giving any cue. The more divergent questions the child generates, the more new videos he can unblock and therefore discover.
    \item \textbf{The evaluation space}: This space contains different domain-knowledge quizzes. The items can be skipped if the child decides that he doesn't know the answer. If, however, the child submits an answer, the agent asks him to report his confidence level in this answer with a 5-Likert scale (from "Super not confident" to "Super confident").
\end{itemize}

\begin{figure}[hp]
\centering
\captionsetup{justification=centering,margin=1cm}
\includegraphics[width=1\linewidth]{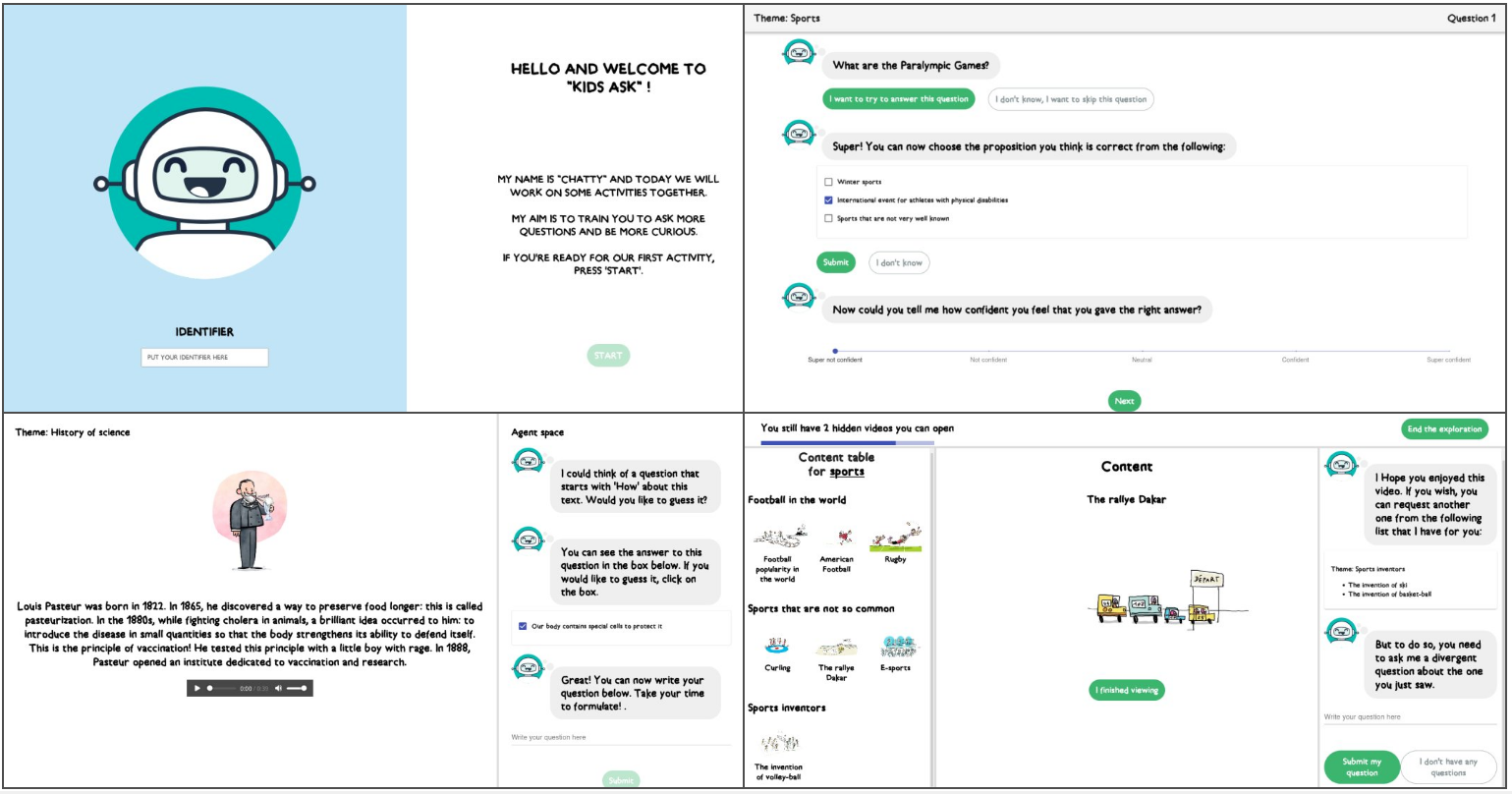}
\caption{\textbf{Workspaces available in "Kids Ask": Homepage (Top left), The question-asking training wrokspace for the experimental condition (Bottom left), The evaluation workspace (Top right) and the exploration workspace (Bottom right)}}
  \label{fig:pres-ka}
\end{figure}

See Figure\ref{fig:pres-ka} for the interfaces of the different workspaces and Section\ref{section: procedure} for bigger snips.

We organized the interaction in the same order for all participants: they all started with a general quiz relating to 6 themes then chose their favourite 2. They then went for the QA training space, followed by the exploration space. They also had a  domain-knowledge evaluation quiz relating to the theme they chose for exploration both before and after this latter.

\subsection{Experimental conditions}
\label{section: conditions}
We had two conditions for this study. The difference between the two conditions resides only in the agent's behavior in the QA training space: For the control group, the agent helps the child only by giving questioning words that could start curiosity-driven questions about the text (example: 'What is'). Whilst for the experimental group, the agent also gives an answer to a possible curiosity-driven question relevant to the text and asks the child to think of a question that leads to this answer (example: the cues 'What is' and 'It is an illness caused by an intestine infection' are meant to lead the child to think of the question 'What is the Cholera disease?'.) In both cases, the child was free to either choose to use these propositions or to think of his own question. See table \ref{tab:dialogue} for explicit examples of the difference in the agent's behavior.

In the exploration and evaluation spaces, the agent exhibited the exact same behavior for both conditions.

\begin{center}
\captionsetup{justification=centering,margin=1cm}
\begin{table}
\begin{tabular}{ | m{12em} | m{7cm} |}
  \hline
  Control condition & \textbf{Agent}: "I could think of a question that starts with \textbf{'What difference'} about this text, try to guess it!.  \linebreak \linebreak \textbf{Agent}: "Remember, the answer to my question is not in the text. Take your time to formulate!" \\ 
  \hline
  Experimental condition & \textbf{Agent}: "I could think of a question that starts with \textbf{'What difference'} about this text.  You can see the answer to it in the box below. If you would like to guess it, click on the box."  \linebreak \linebreak \textbf{Agent:} Renders the checkbox \textbf{'The vaccine prevents the disease, the medication treats it'}. \linebreak \linebreak \textbf{Agent}: "Super, you can now write your question. Take your time to formulate!"  \\ 
  \hline
\end{tabular}
    \caption{\textbf{Illustration of the difference in the agent's dialogue between the two conditions}}
    \label{tab:dialogue}
    \end{table}
\end{center}

\subsection{Design rationale}
\label{chapter:rationale}
For the QA-training space, We chose to focus our training on divergent-thinking questions rather than convergent-thinking ones as the former relate more to epistemic curiosity and involve higher-order thinking and more cognitive processes (as mentioned earlier in section \ref{section: div-ques}). In addition, we choose this design based on Alaimi’s work \cite{Alaimi0}, which had both divergent- and convergent-thinking oriented semantic cues and showed that the agent was actually able to help children generate significantly more divergent questions by using the divergent-thinking oriented semantic cues. Finally, we choose to maintain the linguistic help for both conditions (the questioning word) to help overcome the linguistic difficulties children can face when doing a question-generation task (i.e. not being familiar with the syntactic construction of a question or how to formulate an interrogation), as shown in \cite{Graesser0}. 

Regarding the exploration, the task was designed to investigate whether participants were able to maintain the divergent QA skill that was trained in the previous space (i.e. agents' ability to transfer this skill). We also wanted to investigate if the question-asking skill helped them initiate and maintain autonomous explorations. Furthermore, the data collected in this space also serves us as a behavioral measure of epistemic curiosity. In fact, we use the idea developed in \cite{Gross0} where the authors measure curiosity as the individual’s willingness to ‘work’ for information. Since we constraint children to think of divergent questions if they want to reveal new educational videos, these questions become the effort participants make when the only reward they get in return is the information (i.e. the video) itself. 

Finally, we use the same quiz before and after the exploration phase in order to assess the participants’ learning progress and the impact of the curiosity-driven behaviors on it. The platform's quizzes are skippable and ask children to self-report their confidence levels in their answers. Indeed, research such as in \cite{Graesser0} shows that children do not engage in information-seeking behaviors mainly because they fail to identify their uncertainties. We therefore use the quiz and the confidence-level rating to try to elicit their curiosity by raising their awareness about information they might be missing. In addition, work in \cite{Roebers0} suggests that giving children the possibility to skip a question allows them to think deeper about what they do and do not know and to report low certainty.

\subsection{Design Approach}
We adopted a participatory design approach and included the different stakeholders in our process (i.e., the teachers and the students). To do so, and after developing a first prototype of "Kids Ask", we proceeded to validate it with three primary school teachers. They helped evaluate and propose modifications to the pedagogical approach, the content of the application and the study course (number and length of sessions). 

We then began testing this version with 8 students. We saw difficulties in reading the texts and understanding the task. We then proceeded to change the content to be simpler and added audios for every text in the QA training space to help overcome reading problems. Finally, we tested the new version with 8 other students and saw that they were able to do the task successfully. 

\subsection{Technical implementation}
\begin{figure}[h]
  \centering
  \fbox{
  \includegraphics[width=1\linewidth]{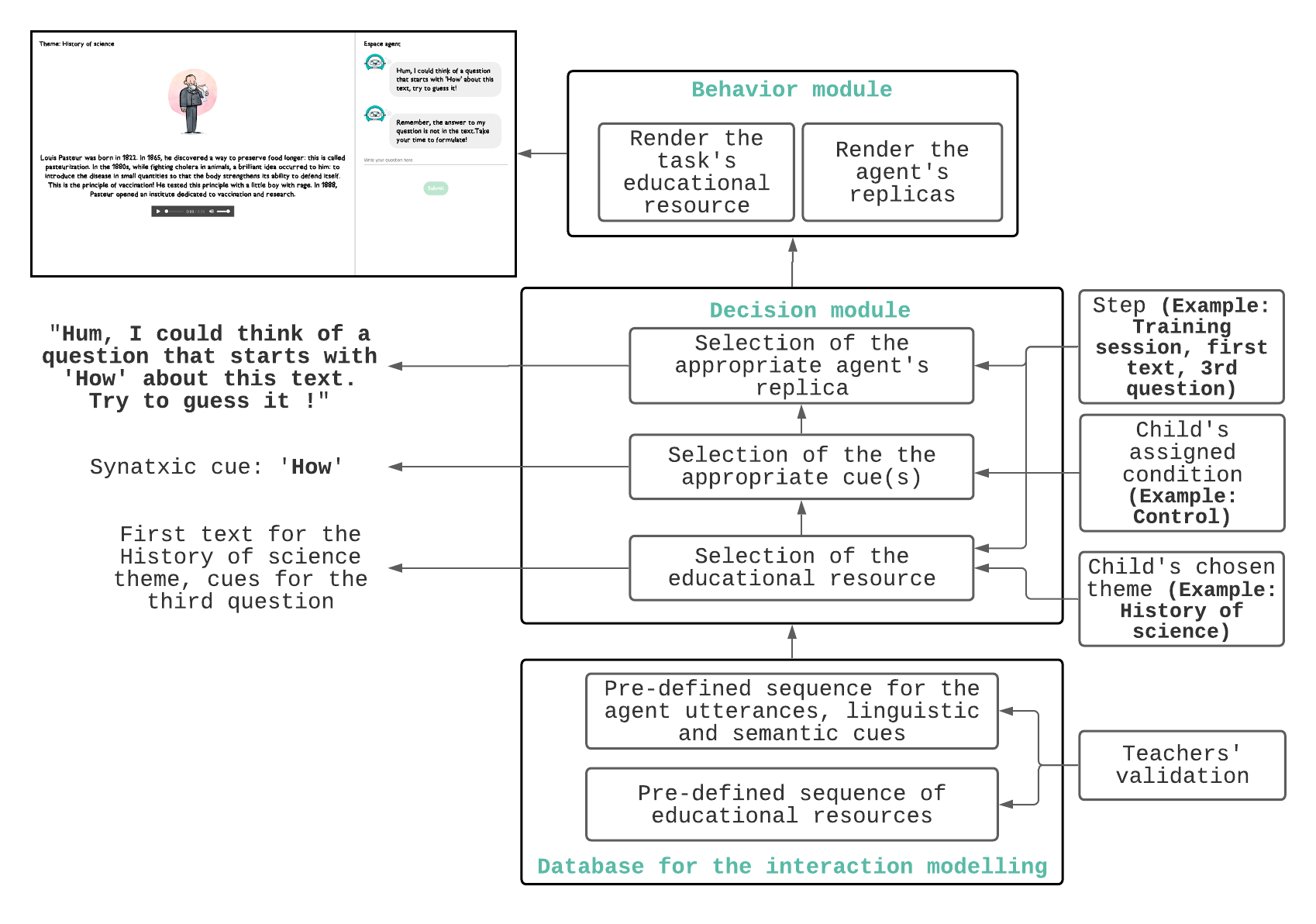}}
  \caption{\textbf{System design}}
  \label{fig:Implementation}
\end{figure}

The interface was programmed in JavaScript using the REACT library and was connected to a RESTful API to publish and retrieve the interaction data. As presented in the diagram below (Figure\ref{fig:Implementation}), the behavior of the agent in terms of selection of the adequate cue(s) to offer was predefined and hand-scripted: it was connected to a database containing the different text resources and every text had a sequence of linguistic and semantic cues linked to it. Depending on the child's condition, the agent's automaton composes the dialogue utterances in order to include the appropriate support. We changed the utterances between the questions to avoid repetition in the agent's dialogue: a replica is not executed if it has been delivered during the previous question \cite{Jones2018}. 

\section{Methodology}
\subsection{Participants}
We recruited 57 CM1 students belonging to four classes from two French primary schools, they were between 9 and 10.5 years old. We were constrained to remove the data for 6 participants who had missing or unusable data (entered incomplete or incomprehensible phrases). This left us with 51 participants that were assigned either to the control group (26 with 12 boys and 14 girls) or the experimental group (25 with 13 boys and 12 girls).

The groups were assigned with a pseudo-randomized method after collecting profile data regarding the age, the device use frequency, the curiosity-related perceptions and the curiosity trait, the reading score and the initial question asking fluency performance (see \ref{section:mesures} for more details of the measures). Thereby, and as shown in Table \ref{Tab:mesures}, we had two balanced groups that were not different in terms of initial profile measures.

\begin{table}
  \begin{tabular}{rclll}
    \toprule
    &Measure&Control group&Experimental group&p-values\\
    \midrule
    \multirow{2}{1em}{}&Age & 9.36 $\pm $ 0.43& 9.31 $\pm $ 0.43&0.66\\
    &Device use frequency & 30.3 $\pm $ 6.28 & 27.24 $\pm $ 7.59&0.12\\ \hline
    \multirow{2}{1em}{}&Curiosity trait & 29.3 $\pm $ 4.3 & 27.12 $\pm $ 4.8&0.1\\
    &Perception of curiosity & 38.8 $\pm $ 8 & 37.5 $\pm $ 6.58&0.53\\ \hline
    \multirow{2}{1em}{}&Reading ability & 280.03 $\pm $ 50.5 & 272.94 $\pm $ 96.6&0.75\\
    &QA fluency & 9.61 $\pm $ 7.38& 8.24 $\pm $ 5.56&0.46\\
  \bottomrule
\end{tabular}
  \caption{\textbf{Profile measures for the experimental and control conditions}}
    \label{Tab:mesures}
\end{table}

\subsection{Procedure}
\label{section: procedure}
\begin{figure}[h]
  \centering
  \fbox{
  \includegraphics[width=1\linewidth]{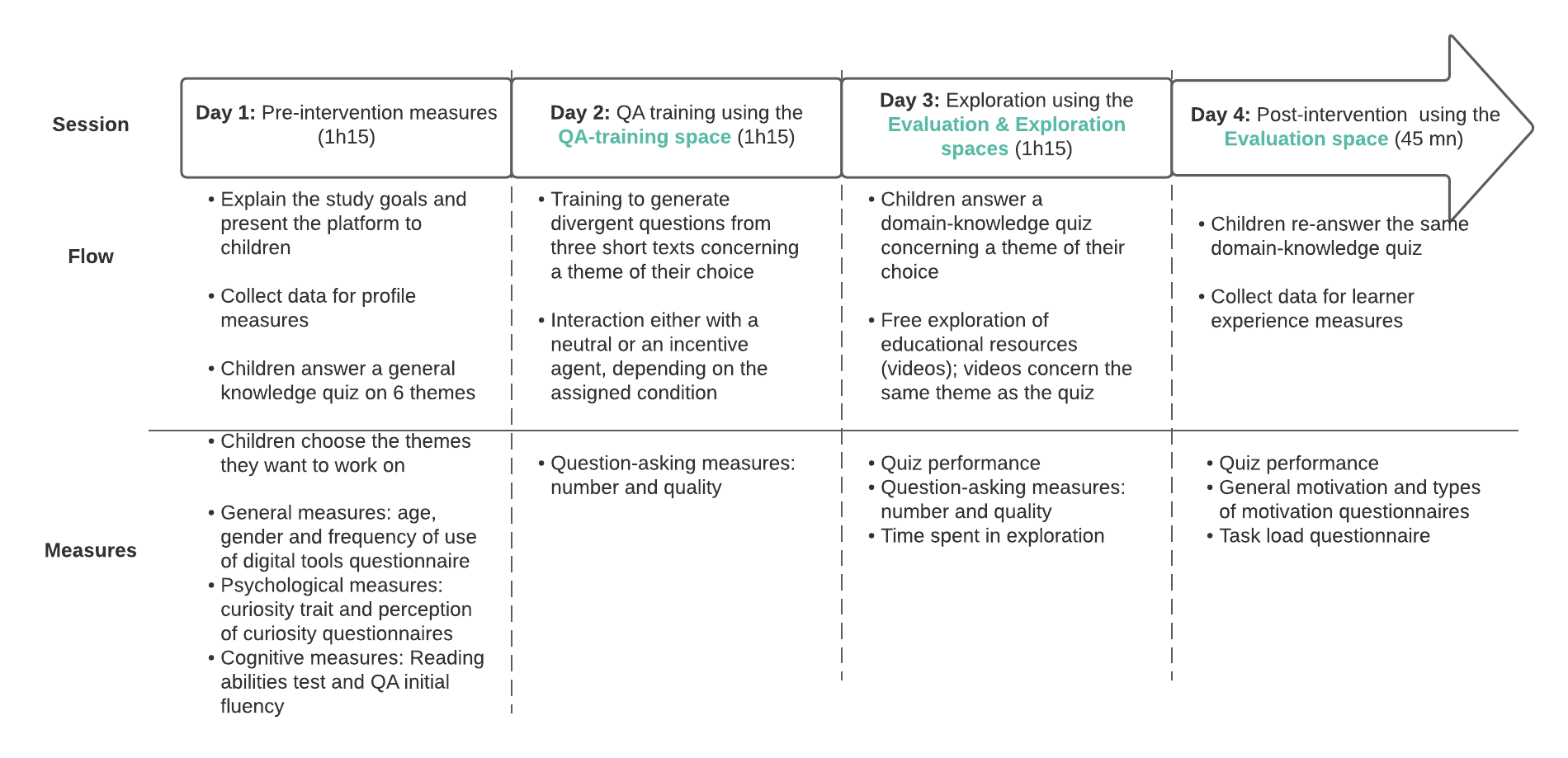}}
  \caption{\textbf{Study timeline and measures}}
  \label{fig:timeline}
\end{figure}

The experiment consisted of four sessions of 1h15 within the same week: One for collecting the profile measures mentioned above, one for the divergent-QA training, one for the exploration and one for the post-intervention assessment. See study timeline in Figure \ref{fig:timeline} and details of the different measures in the data collection paragraph \ref{section:mesures}. 

Our study was approved by the institute's ethics committee (certificate n°2019-23) and only started after having all participants parents' signed consents.

\begin{figure}[h]
  \centering
  \fbox{
  \includegraphics[width=1\linewidth]{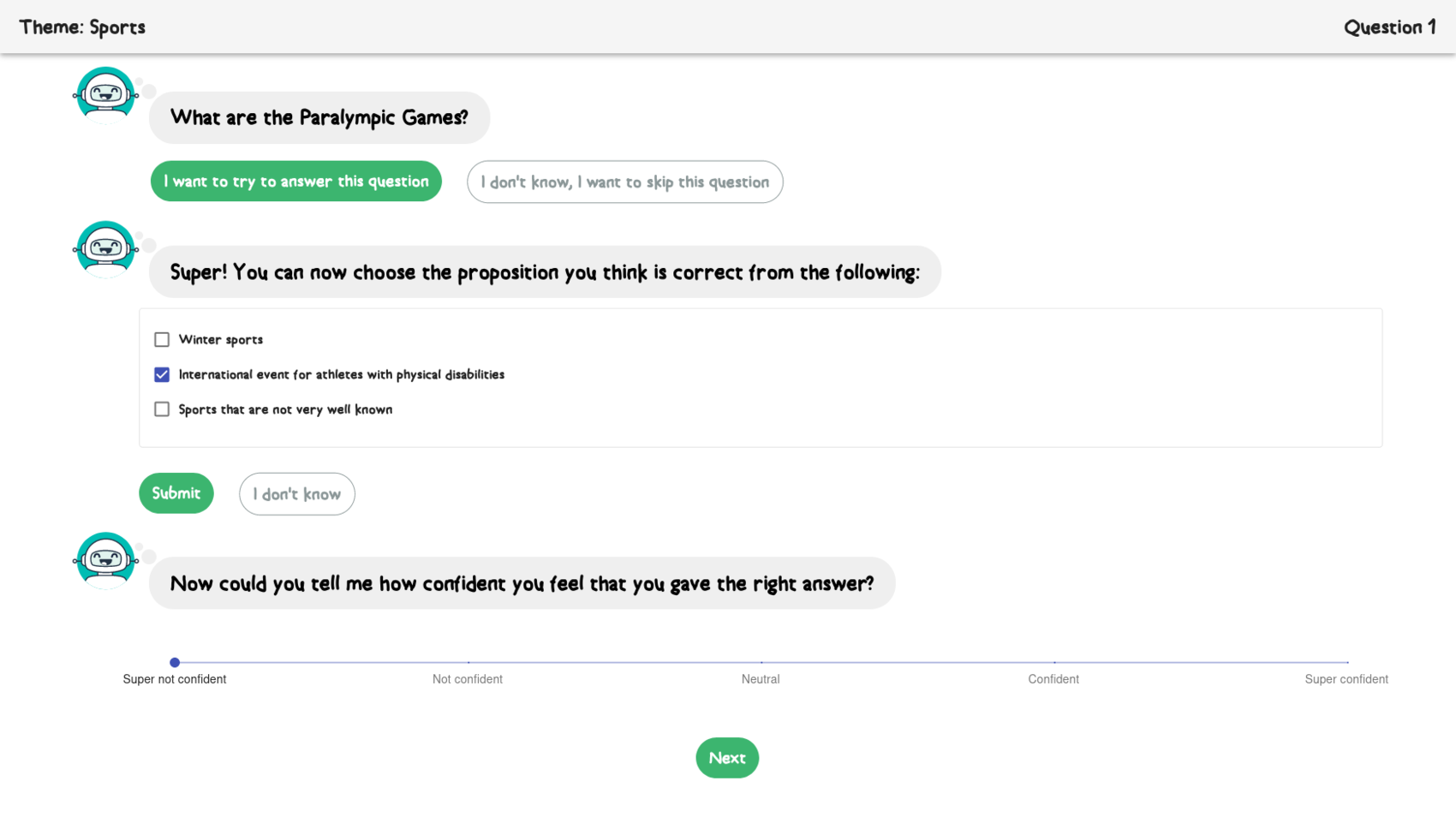}}
  \caption{\textbf{Interface for the evaluation workspace}}
  \label{fig:quiz}
\end{figure}

\textbf{Session 1} During the first session, we presented the study to the participants and collected the age and gender data. We also administered the questionnaires concerning the curiosity trait, the perception of curiosity and device use frequency. We then run the reading abilities test and the question-asking fluency test (see more details of the tests in the chapter \ref{section:mesures}). We also took time to explain what divergent questioning means and highlighted the difference between divergent and convergent questions. Participants then went to the evaluation space of "Kids Ask" and began the interaction with a general quiz (pre-test) relating to six themes. For each question, children had the choice to either skip the item, by clicking on the 'I don't know, I want to skip this question' button, or answer it. If they do choose to answer, they are asked to report their confidence level in their answer: they had a 5-Likert scale from 'Super not confident' to 'Super confident' (see Figure \ref{fig:quiz}).

Once this pre-test completed, children got to choose their two favourite topics and were told that they will work on these latter during the upcoming sessions. 

\textbf{Session 2} During the QA training session, participants were asked to read short texts relating to one theme they chose. They then started working on generating divergent-thinking questions relating to these texts using the cues provided by the agent. They also had audio players for each text to help them if they had reading difficulties. The texts were selected from online resources and children magazines (Sciences et vie Junior, Quelle Histoire and Questions? Réponses!) and were edited in order for them all to have six sentences and the same number of words. The average number of words per text was of 109.

\begin{figure}[h]
  \centering
  \fbox{
  \includegraphics[width=1\linewidth]{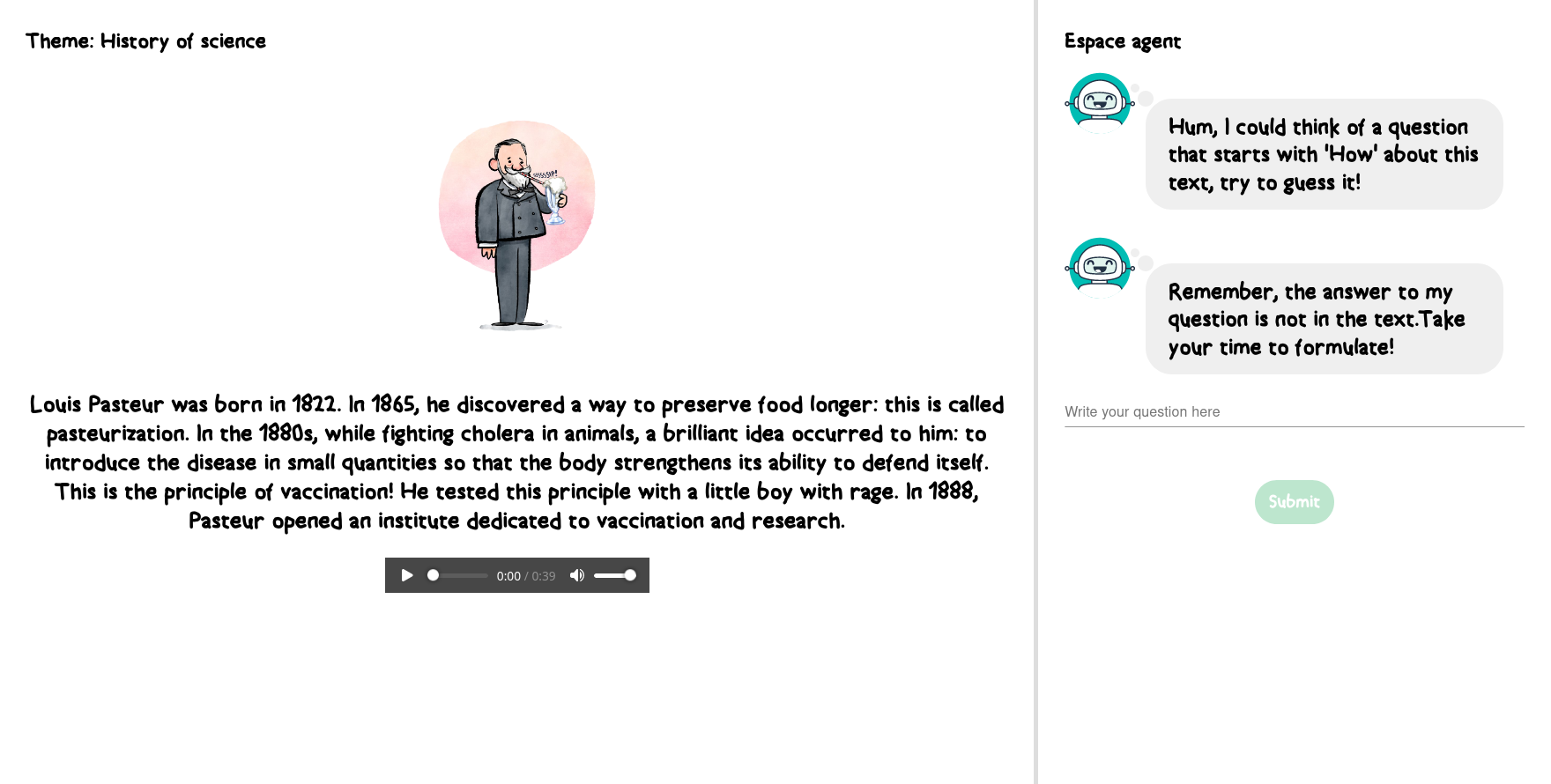}}
  \caption{\textbf{Agent behavior for the control condition during the QA-training session}}
  \label{fig:qa-c}
\end{figure}

\begin{figure}[h]
  \centering
  \captionsetup{justification=centering,margin=1cm}
  \fbox{
  \includegraphics[width=1\linewidth]{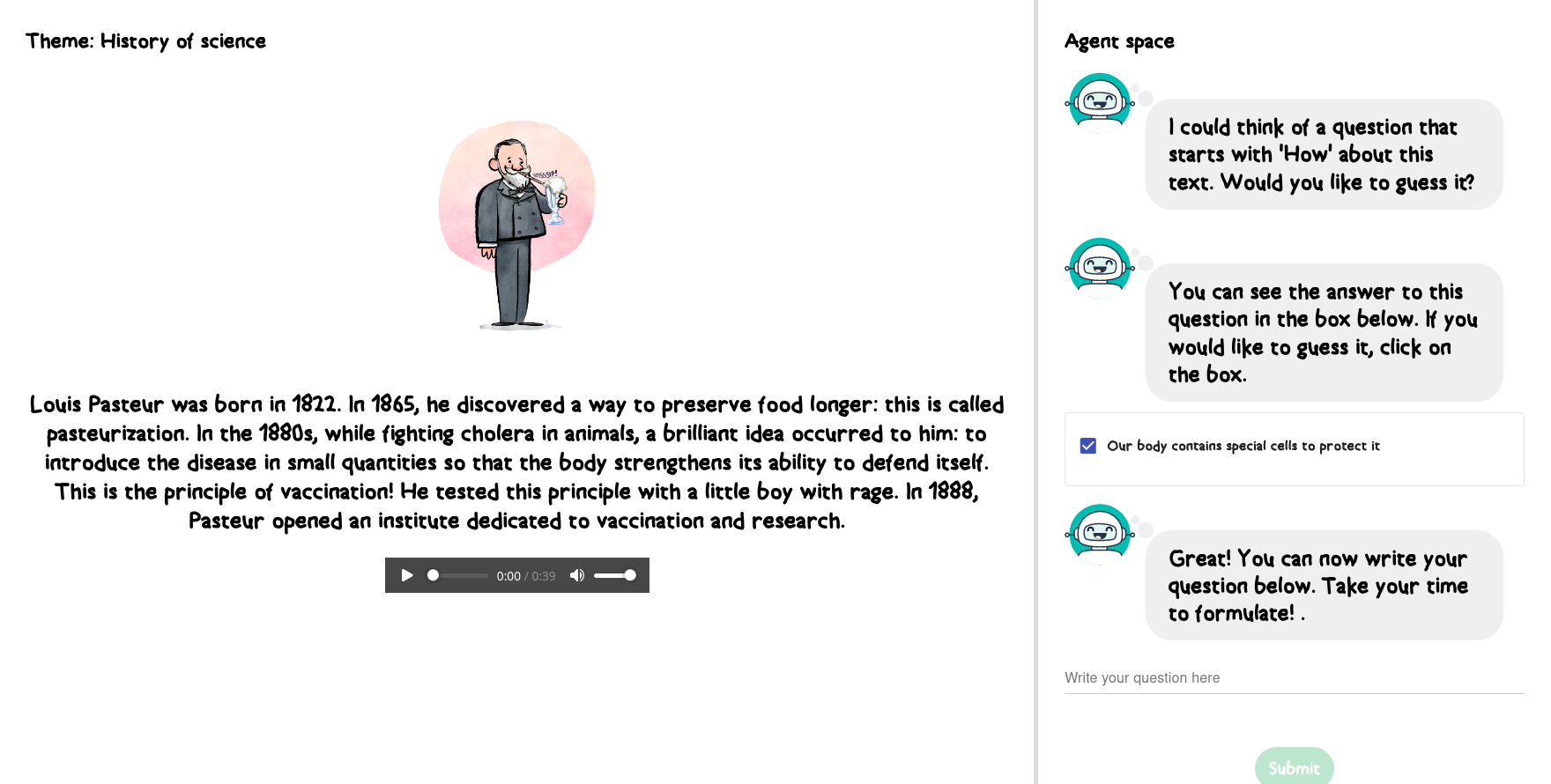}}
  \caption{\textbf{Agent behavior for the experimental condition during the QA-training session}}
  \label{fig:qa-e}
\end{figure}

To begin working on a text, participants were asked to read or listen to it and then click on the ‘I finished reading’ button once they understand it. This button enables the ‘discussion’ with the agent, in the agent's space on the right section of the screen. As explained above, the agent helps the child generate divergent-thinking questions about the text by giving either linguistic support (control group) or linguistic plus divergent-thinking semantic support (experimental group). See Figures \ref{fig:qa-c} and \ref{fig:qa-e} for the difference between the conditions. 

During this session, children had to process three texts and generate six questions per text with the agent, making a total of 18 questions. They were not restricted in terms of time, apart from the session length. They had the application running on tablets and worked individually.


\textbf{Session 3} Participants started with a quiz concerning the second theme they chose and then moved on to the exploration phase. During the exploration, they were encouraged to navigate autonomously amongst several educational videos but were required to use their divergent-questioning skills to do so. The videos were all taken from the same french website 1 jour, 1 question. They were all between 1mn20 and 1mn30 of length and contained the same amount of information and the same number of sentences and drawings.

\begin{figure}[h]
     \centering
     \fbox{
     \includegraphics[width=1\linewidth]{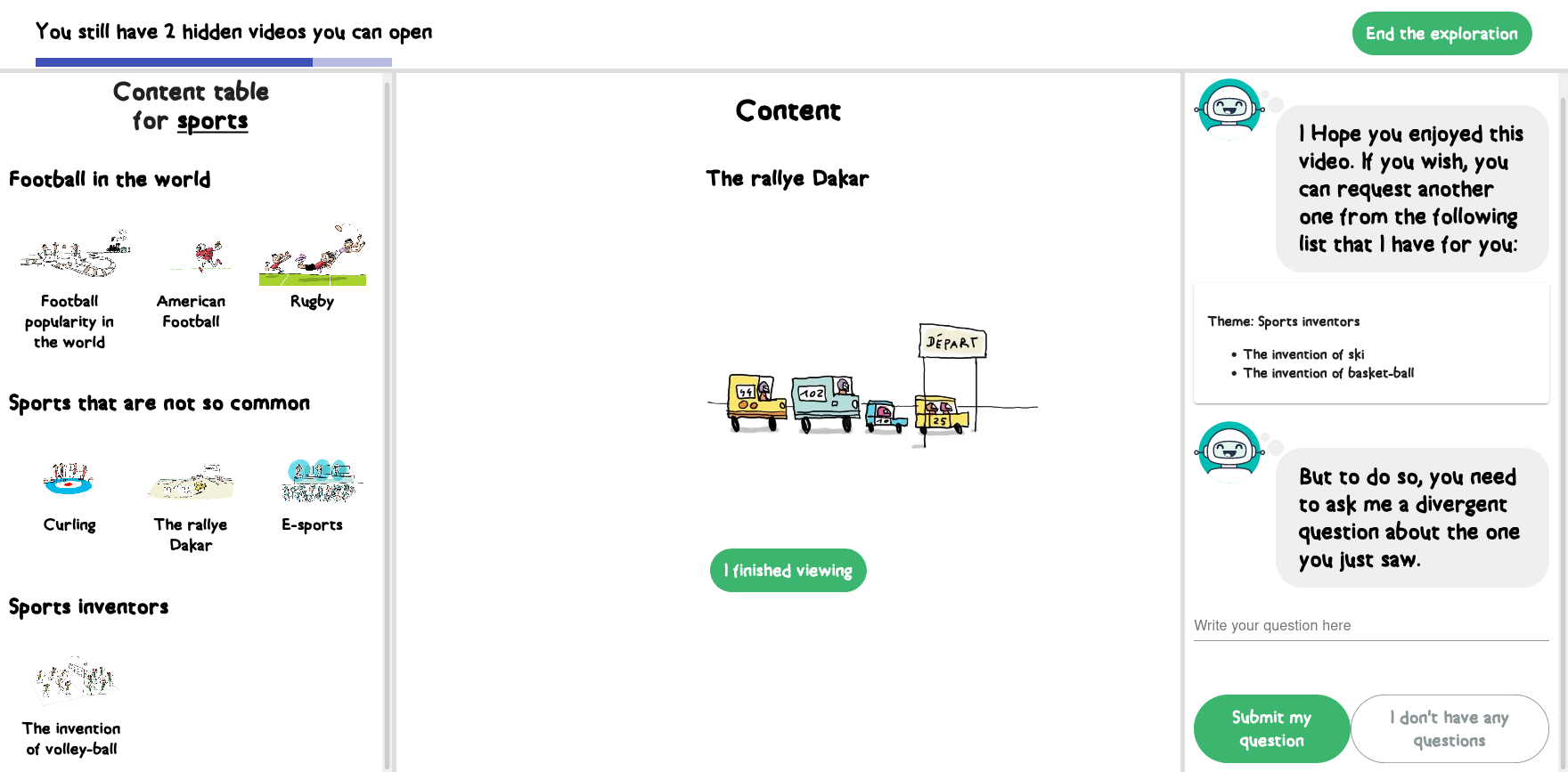}}
     \caption{\textbf{Agent behavior after finishing viewing a video during the exploration}}\label{Fig:explopre}
   \end{figure}

\begin{figure}[h]
  \centering
  \fbox{
  \includegraphics[width=1\linewidth]{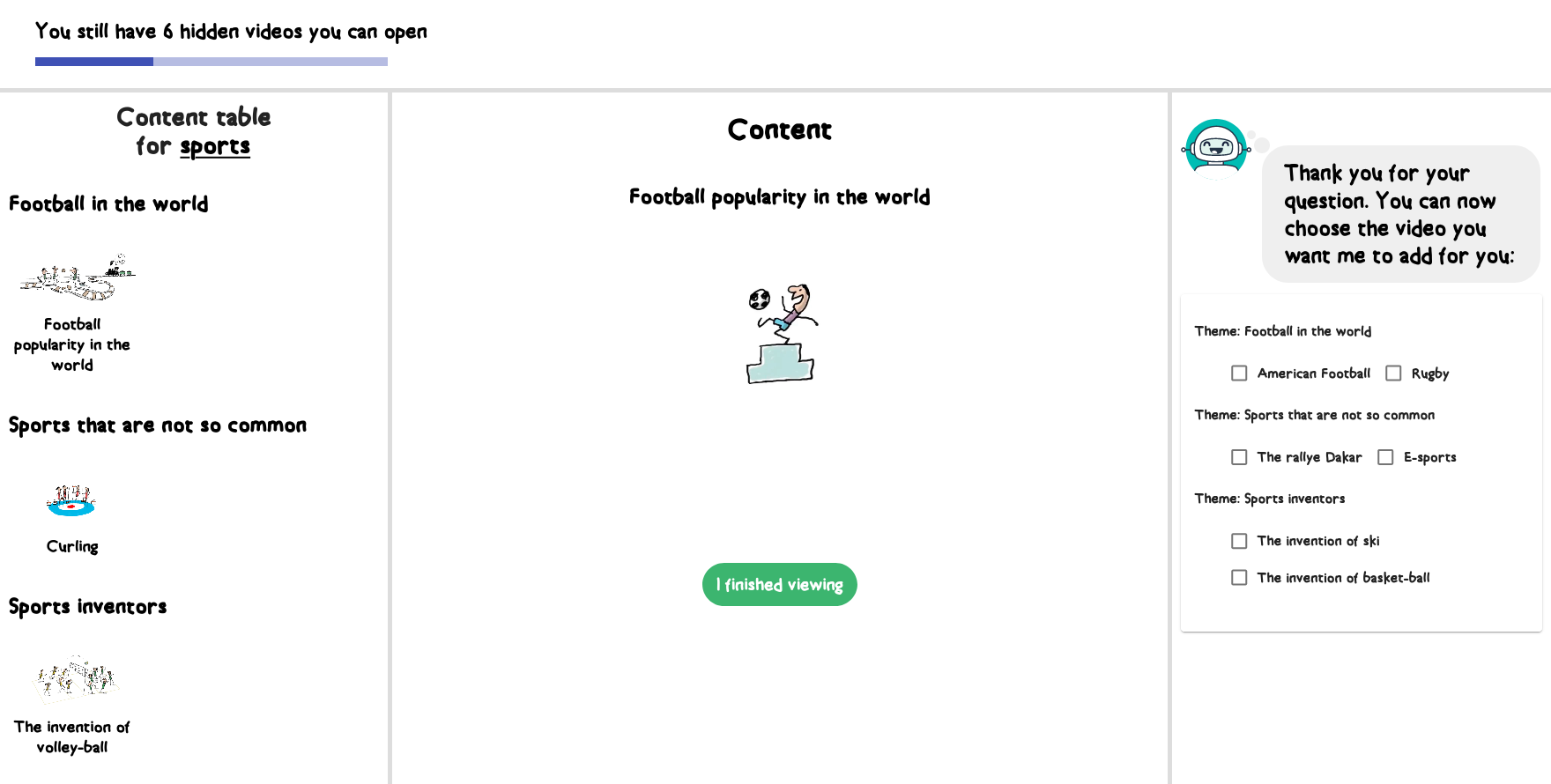}}
  \caption{\textbf{Adding the new video "American football" to the content table}}
  \label{fig:explo-choice}
\end{figure}

\begin{figure}[h]
  \centering
  \fbox{
  \includegraphics[width=1\linewidth]{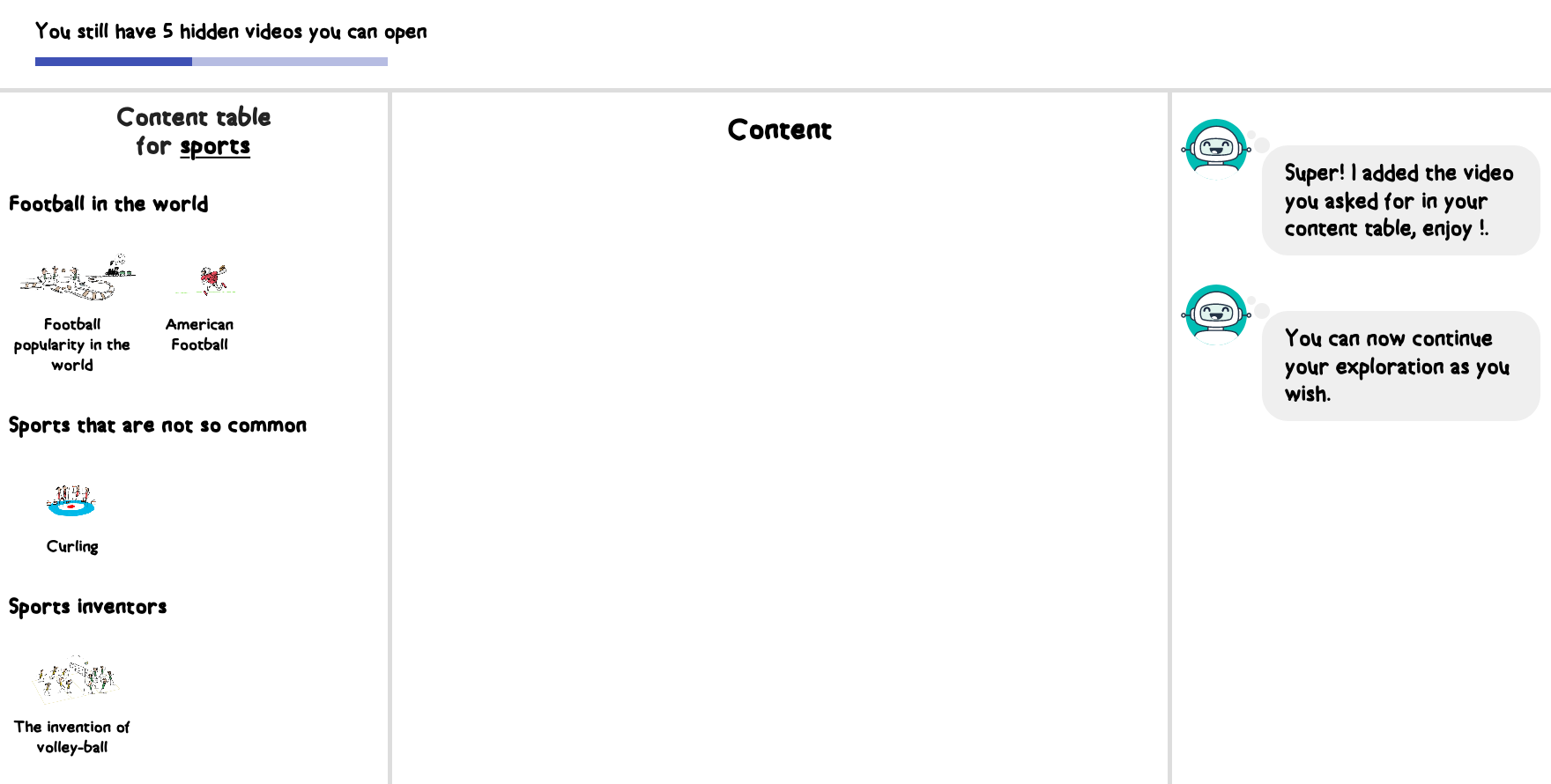}}
  \caption{\textbf{Agent behavior after submitting a question during exploration}}
  \label{fig:explo-q}
\end{figure}

At the beginning of exploration, participants had 3 educational videos related to their second theme in their 'Content table' section, and a progression bar suggesting that there are 6 more hidden ones. When a participant selects a video from these three, it appears in the central 'Content' section and he is able to play it. He is then asked to click on the ‘I finished viewing’ button once he finishes playing it. This button makes the agent reappear in the right section of the page. It now shows the participant a list of the remaining hidden videos he can unblock for him (see Figure \ref{Fig:explopre}). It then informs him that in order to choose a new video to add to his content table, he is required to enter a divergent question about the one he just saw. If the child decides that he has no questions, the agent has no actions and the space remains unchanged. If, on the other hand, the child submits a question, he can choose a new video to open from the list. Once chosen, the video is added to his pedagogical content and becomes available for watching. The progress bar continuously indicates the remaining number of locked videos (See Figures \ref{fig:explo-choice} and \ref{fig:explo-q} for the agent's behavior before and after the participant enters a question and chooses a new video). 

There is a total of three initially-unlocked videos (children could navigate between them without generating questions), six new videos to unlock and 9 questions to generate over this session. Children were able to deliberately stop this session by clicking on the ‘I finished exploring’ button that appears on the top right of the screen, after opening at least three videos.


\textbf{Session 4} Participants re-answered the same quiz they had during the pre-exploration. They also answered the post-intervention surveys for the general motivation, types of motivation and the task load (see measures timeline in \ref{fig:timeline} and further descriptions in the "Data collection and instruments" chapter \ref{section:mesures}).

\subsection{Data collection and measures}
\label{section:mesures}
\subsubsection{Profile Measures}
This includes general measures: age, gender and the technology frequency use, psychological measures: curiosity trait (\cite{Litman0}) and perception of curiosity using the CIAC questionnaire (\cite{Post0}). And finally, two cognitive measures: the reading abilities using the standardized test in \cite{Alouette} and the question-asking fluency: this is an offline test where we give participants a short text to read then ask them to write as many questions about it as they can, within 2 minutes. The QA fluency measure consists of computing the sum of the scores of the questions generated during these 2 minutes, using the quotation grid that assesses the questions quality, described in Section\ref{section:q-score}.

\subsubsection{Behavioral curiosity measures}
\begin{itemize}
\item
\emph{Percentage of divergent questions asked.}
This measure consists of counting the total number of the correct questions generated during each session and computing the percentage of the curiosity-driven, i.e. the divergent, questions amongst them. We did not consider the correspondence between the question entered and
the cues given by the agent, so children were free to imagine another
question if they wanted to, as long as it is still related to the resource. If
a question is repeated, it is only taken into account once.

A question is considered correct and taken into account if:
\begin{itemize}
    \item It is a question --- and not a statement.
    \item It is related to the educational resource (text or video).
    \item It is not repeated more than one time.
    \end{itemize}
\textbf{Example:} For the text about Louis Pasteur (available in Figures \ref{fig:qa-c} or \ref{fig:qa-e}), a   linguistic cue ‘\textbf{What are the other}’ and a semantic cue “\textbf{Fermentation}”, we accepted  questions such as: \begin{itemize}
    \item “What are the other possible ways to preserve food?” : used the agent propositions and is related to the text.
    \item “What are the other ways to protect our body?” : did not use the agent semantic proposition but is still related to the text.
    \item “How is the vaccine different from medication?” : did not use both of the agent propositions but is still related to the text.
\end{itemize}  
However, data such as the following was not accepted:
\begin{itemize}
    \item “It is another way to preserve food“ : this is a statement and not a question.
    \item "What are the other parts of a robot?" : this is not related to the Louis Pasteur text.
    \item "What are the other characters of the Simpsons?" : this is not a serious attempt.
\end{itemize}. 

On another hand, a question was considered to be divergent if its answer is not explicitly stated in the educational resource (text or video) in question.

\textbf{Example:} The question “What are the other possible ways to preserve food?” is considered to be divergent, whereas "What are the other elements of the Pasteurisation process?" is considered convergent as the answer to it is explicitly stated in the text.

This annotation process was performed by the two researchers who led the experiments in schools. All data was anonymized: coders could only see the identifiers that children were given randomly at the beginning of the intervention. The two researchers annotated 65\% of the data generated during the the different phases. The inter-rater reliability was of 88,1\%, with an agreement percentage of 90\%. The rest of the data was only coded by the first author.

\item
\emph{Quality of the questions generated}
\label{section:q-score}
In scoring the quality of the questions generated, we used the grid that emerged from the classification in \cite{Ascher0, Chouinard0} and computed a question's score as the sum of the following criteria: 
\begin{itemize}
    \item One point if the question is high-level: the answer to this question is not a simple fact (example: 'How big is a dinosaur?') but requires to explain a mechanism, a relationship etc (example: 'Why were dinosaurs so big?').
    \item From 1 to 4 points, based on the syntactic construction of the question : \begin{itemize}
        \item 1 point for a ‘closed’ or declarative question (example: "Dinosaurs were big?").
        \item 2  points for questions with questioning words in the middle of the sentence (example: "The dinosaurs were how big?").
        \item 3 points for a question without an interrogative formulation (example: "Why the dinosaurs are big ?").
        \item 4 points for a questioning word in the beginning of the sentence that has interrogative syntax (example: "Why are dinosaurs big?").
    \end{itemize}  
    \item From 1 to 3 points, based on the use of questioning words : \begin{itemize}
        \item 1 point for a declarative question i.e., with no questioning word (example: "Dinosaurs were big?").
        \item 2 points for questions with ‘Is/Are’ (example: "Are dinosaurs big?").
        \item 3 points for the use of proper questioning words (example: "How big were the dinosaurs?").
    \end{itemize}  
\end{itemize} 

Ultimately, questions had scores from 2 to 8 points. For each participant, we calculated an average score for all the questions generated during each session.

Again, the same two raters coded 65\% of the questions generated following this grid and had an inter-rater reliability of 79.6\%, with an agreement percentage of 84.8\%. The rest was coded by the first rater.

\item
\emph{Time spent in exploration.}
This metric reflects the number of videos explored. We consider this as a curiosity measure as children were told that they are able to end the session if they feel that they are not curious about discovering the new resources anymore.

\end{itemize}

\subsubsection{Domain-knowledge learning progress.}
Participants had to answer the same domain-knowledge quiz before and after the exploration phase; the maximum score is of 9 points. The items were like the following:
\begin{itemize}
    \item 1/3 of the questions had their explicit answers in the videos.
    \item 1/3 of the questions required the participant to link information from different videos to find an answer.
    \item 1/3 of the questions did not relate to the theme chosen for the exploration. These are our control questions to help us investigate if the learning progress made is actually related to the exploratory behavior.
\end{itemize}

The learning progress was then computed in order to evaluate the participant’s progress with respect to the maximum learning gain achieved: 
\begin{equation}
 learning\ progress = \frac{Score_{postExploration}-Score_{preExploration}}{Score_{Max}}
\end{equation}
With: $Score_{preExploration}$ and $Score_{postExploration}$ being, respectively, the scores during the pre- and post- exploration quizzes. And, $Score_{Max}$ being the highest score achieved in the post-exploration quiz.

\subsubsection{Learner experience measures}
\begin{itemize}
    \item 
\emph{Motivation measures.}
The participants’ motivation to use "Kids Ask" was assessed via two questionnaires: the general motivation scale \cite{Cordova0} and Vallerand's types of motivation scale \cite{vallerand0}. 

The general motivation scale was used to investigate the potential short-term motivation. It contains one sub-scale for evaluating participants' satisfaction with the platform, one sub-scale for evaluating their perceived competence and one sub-scale for assessing the degree of preference with respect to a favourite school activity.
The items were answered either with a 6- or 7-Likert scale, with the maximum score being 54.

On another hand, Vallerand's scale was used to probe intrinsic and extrinsic motivational mechanisms in the educational settings we had. It is composed of three sub-scales that differentiate: intrinsic motivation (possible scores from 0 to 9 points), extrinsic motivation (possible scores from 0 to 9 points) and amotivation (possible scores from 0 to 3 points).
All questionnaire items are yes or no questions.

\item
\emph{Task load.}
The participants' behavior was also evaluated in terms of the subjective workload they experienced during the intervention. For this, we used the NASA-TLX workload multi-dimensional scale developed in \cite{Nasa}. Information about the intensity of six workload-related factors are used in order to estimate a reliable measure of workload: mental demand, physical demand, temporal demand, performance, effort and frustration. 
\end{itemize}

\subsection{Research questions and hypotheses}
The main goal of this research is to design a conversational agent that can train children’s divergent-thinking question-asking skills and foster the use of this skill to maintain autonomous exploratory behaviors that lead to domain-knowledge learning progress. We hypothesize that:
\begin{itemize}
    \item The incentive agent, compared to the neutral agent, increases children’s ability to generate divergent-thinking questions.
    \item The incentive agent, compared to the neutral agent, increases children's ability to maintain autonomous explorations.
     \item The incentive agent, compared to the neutral agent, leads to stronger domain-knowledge learning progress.
     \item Differences in the achieved domain-knowledge learning progress can be explained by differences in curiosity-driven behaviors.
     \item The incentive agent, compared to the neutral agent, will be associated with a better learning experience and a higher intrinsic motivation.
\end{itemize}

\section{Results}
\begin{figure}[hp]
\centering
\captionsetup{justification=centering,margin=1cm}
\fbox{
\includegraphics[width=.5\linewidth]{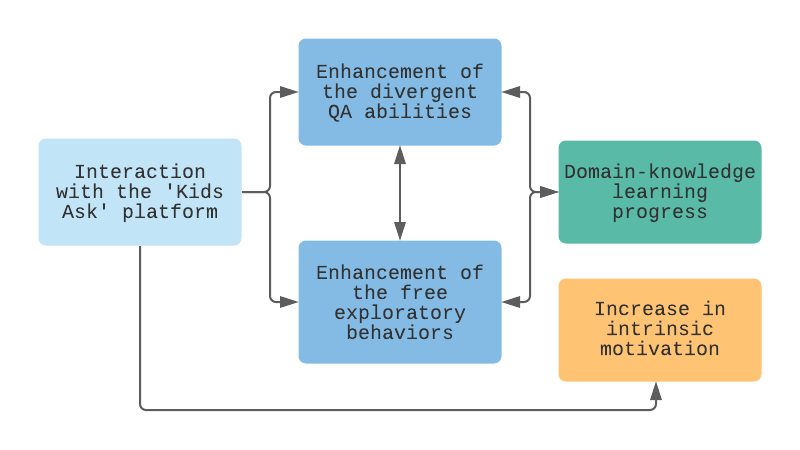}}
\caption{\textbf{Relationships to investigate with the "Kids Ask" platform}}
\label{fig:res-patt}
\end{figure}

To address our research questions, we conduct the data analysis so that we are able to distinguish the effect of interacting with the two agents on children's divergent question-asking abilities. We also analyze if children were able to maintain this skill in the absence of the agent support and use it to lead autonomous exploratory behaviors. On a further step, we investigate the effect of these curiosity-driven behaviors on the domain-knowledge learning progress. And finally, we look into the learners’ experience with the platform in terms of cognitive load and and motivation of use (Figure\ref{fig:res-patt}). 

\subsection{Behavioral curiosity measures}
\subsubsection{Divergent question-asking performance and effect of the training}
\begin{figure}%
    \centering
    \subfloat[\centering \textbf{Participants with the incentive agent generated significantly more divergent questions. Their performance did not drop significantly during the second session, revealing a stronger efficiency for this agent in transferring the divergent QA skill.}]{{\fbox{\includegraphics[width=7cm]{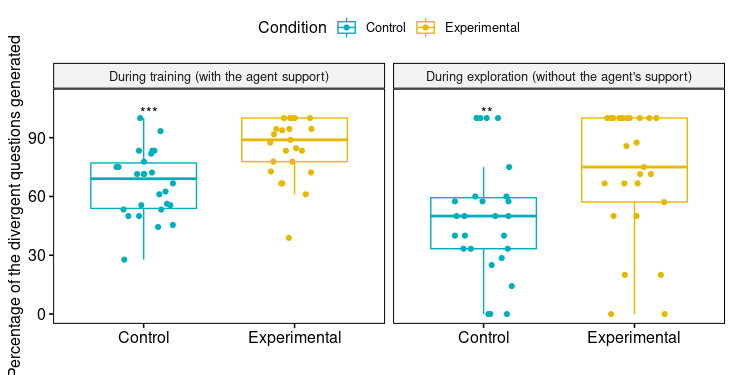}}} 
    }%
    \subfloat[\centering \textbf{Participants with the incentive agent generated significantly better divergent questions. Their performance did not drop significantly during the second session, revealing a stronger efficiency for this agent in transferring the divergent QA skill.}]{{\fbox{\includegraphics[width=7cm]{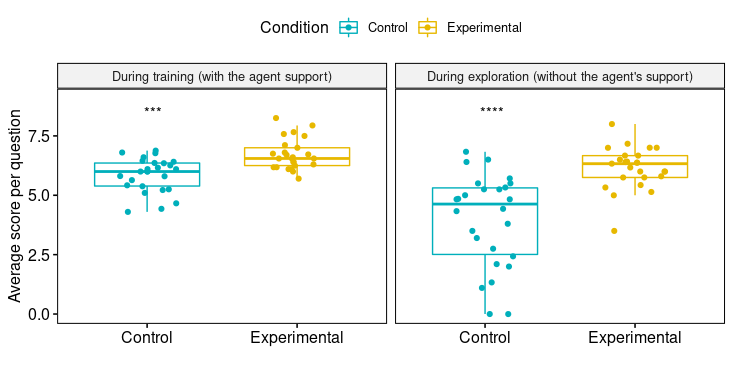} }}}%
    \caption{Question-asking measures during the two phases for the two groups}%
    \label{fig:res-qa}%
\end{figure}

In analyzing the percentage of divergent questions generated, we used a two-way mixed ANOVA test, followed by Post-hoc tests with the Benferroni procedure. Our aim being to evaluate if the performance was different between the two conditions, between the two sessions and, finally, if it depended on the agent's behavior in the same way during both sessions. Results showed a statistically-significant effect of the condition nature on the percentage of divergent questions generated: F(1,49)=17.87; p=0.0001); see sub-figure (a) in Figure\ref{fig:res-qa}. On another hand, the effect of session nature on the divergent QA percentage was also significant (F(1,49)=10.81; p=0.002). The post-hoc tests investigating the simple main effect of the session for each condition level revealed a significant interaction only for the control group: F(1,49)=7.22; p=0.01; the performance dropped drastically. This was not the case for the experimental group: F(1,49)=3.61; p=0.07 where the performance tended to stay stable.
Finally, and as expected, the two-way interaction between the condition and the session on the performance was not significant (F(1,49)=0.19; p=0.67) suggesting that the divergent QA percentage depended on the agent's behavior in the same way during both sessions. 

On a final note, in terms of quality of the questions generated, we used the grid described in section \ref{section:q-score} and calculated an average score per participant and per session. Similar to the previous analysis, results of the two-way mixed ANOVA showed that the assigned condition affected the average score of the questions: F(1,49)=37.48 and p$<$0.0001; see sub-figure (b) in Figure\ref{fig:res-qa}. The output of our test also showed a significant two-way interaction between the condition and the session on the average score of the questions (F(1,49)=9.5; p$<$0.0001) indicating that the positive impact of the sessions on the questions quality depended on the condition nature.

Before running the tests above, assumptions about normality, absence of outliers and homogeneity of variance were verified.

Taken together, these results support the hypotheses that our incentive agent was more successful in training children's faculties to generate divergent questions both in terms of quantity and quality. It was also more successful in facilitating the transfer of this strategy and its use in new contexts, where no cues are given to help.

\subsubsection{Exploratory behavior and effect of the training}
\begin{figure}[h]
  \centering
     \captionsetup{justification=centering,margin=2cm}
  \fbox{
  \includegraphics[width=.5\linewidth]{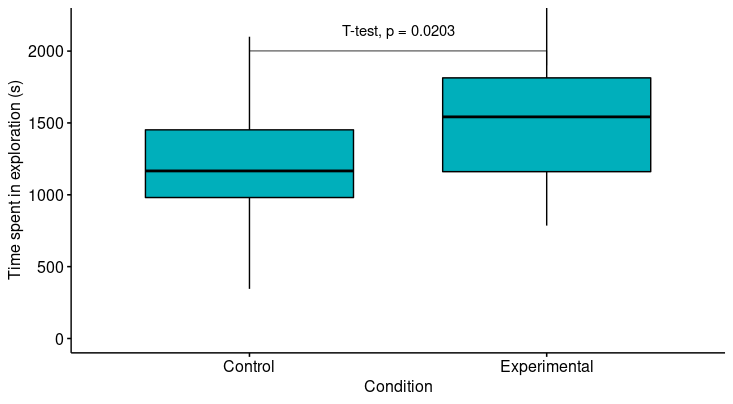}}
  \caption{\textbf{Participants with the incentive agent spent significantly more time exploring the educational resources}}
  \label{fig:temps-explo}
\end{figure}

This metric reflects the number of resources that children were interested in seeing during their exploration. As shown in Figure\ref{fig:temps-explo}, children who interacted with the incentive agent ended up spending more time exploring the educational resources than those who interacted with the neutral agent: m=1222.95 s, SD= 439.76 for the control group and m= 1593.15 s, SD= 634.81 for the experimental; with p-value = 0.02 and an effect size (Cohen's d) of 0.68.

In analyzing the role of the training session on this difference, we run an ANCOVA test between the two conditions, with the time spent in exploration as a dependant variable and the divergent QA performance during the training as a co-variate. Results showed, indeed, a significant interaction: F(1,48)=4.1 and p=0.04. Before running the test, hypotheses about linearity between the response variable and the co-variate, normality and absence of outliers were verified.

Together with the results of the previous section, we can suggest that our incentive training was more successful in enhancing divergent QA faculties that, in their turn, had a significant positive effect on children's ability to lead autonomous explorations.

\subsection{Domain-knowledge learning progress and effect of the divergent question-asking performance}
In analyzing the two groups' learning progress, we compare participants' performances in the domain-knowledge quizzes before and after the exploration. It is to be noted that children's scores were initially similar for the two conditions, meaning that they both had the same space to achieve progress: m=2.73$\pm$ 2.34 for the control group, m=2.32$\pm$ 2.91 for the experimental group (maximum score =9) and p-value= 0.58.

\begin{figure}%
    \centering
    \subfloat[\centering \textbf{Participants with the incentive agent had a significantly higher learning progress even though they had similar initial scores}]{{ \fbox{\includegraphics[width=6cm]{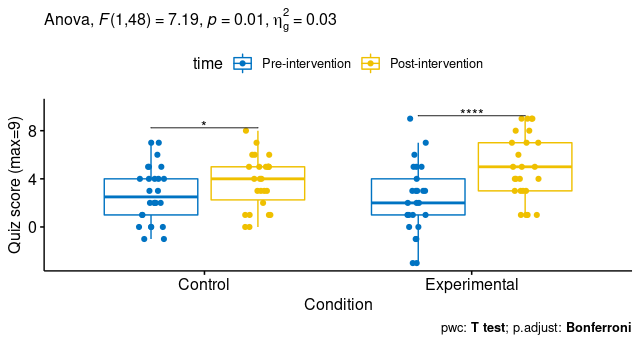} }}}%
    \subfloat[\centering \textbf{The difference in learning progress is explained by the differences in the divergent QA performances} ]{{\fbox{\includegraphics[width=6cm]{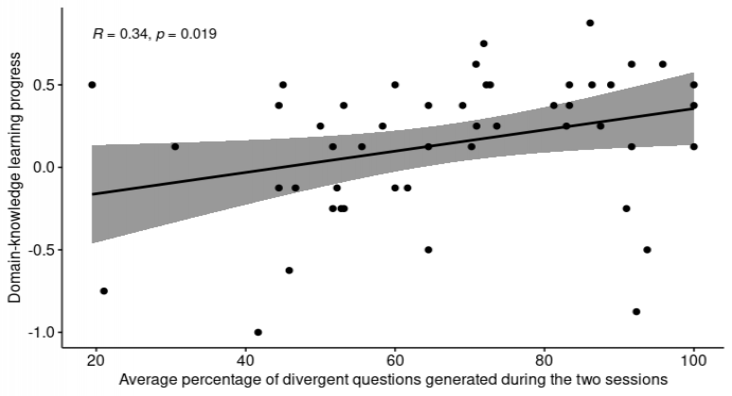} }}}%
    \caption{Domain-knowledge progress and effect of the question-asking skills}%
    \label{fig:res-lp}%
\end{figure}

As shown in Figure\ref{fig:res-lp}, the experimental group had a higher learning progress : m=0.25$\pm$ 0.41 compared to m=-0.009$\pm$ 0.45 for the control group; p-value= 0.04. Furthermore, we conducted a two-way repeated measures ANOVA test to see if the difference between the initial and final performances was significantly different between the two conditions. The interaction between time and the nature of the condition on the domain knowledge score came out statistically significant: F(1,48)=6.42 and p=0.015. Hypotheses concerning the absence of outliers and normality of the learning progress variable were verified pre running the test.

Finally, in analyzing the the role of the divergent QA performance on learning, we run an ANCOVA test between the conditions with the learning progress as a dependant variable and the average divergent QA performance for both sessions as a co-variate. The interaction was statistically significant with F(1,48)=7.19; p-value=0.01. The ANCOVA conditions were also verified.

With these results, we saw that the two groups were different regarding the learning progress that they achieved autonomously, and that this difference was mediated by how much divergent questions they were able to generate.

\subsection{Learner experience measures}
\subsubsection{Motivation measures}
\begin{figure}%
    \centering
    \subfloat[\centering \textbf{Participants were significantly more intrinsically than extrinsically motivated during the two sessions but did not differ between the two groups}]{{ \fbox{\includegraphics[width=6cm]{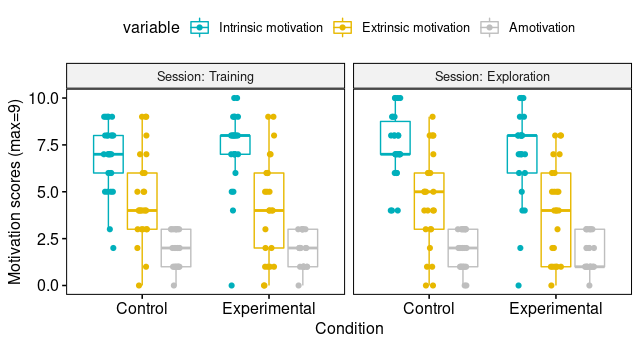} }}}%
    \subfloat[\centering \textbf{The tasks load tended to be slightly lower for participants with the incentive agent} ]{{\fbox{\includegraphics[width=6cm]{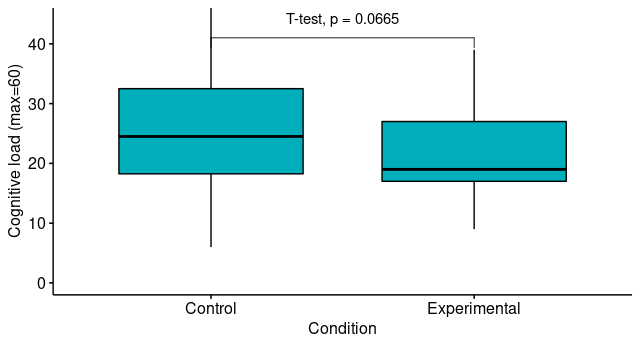} }}}%
    \caption{Learner experience measures}%
    \label{fig:res-ui}%
\end{figure}

The scores of general motivation did not differ between the two conditions (p=0.54) but remained high throughout the two sessions (m=39.3 for the experimental and m=37.74 for the control; max score =54). The difference between the groups in terms of intrinsic motivation was not significant either. However, and as shown in the sub-figure (a) in Figure\ref{fig:res-ui}, participants were significantly more intrinsically than extrinsically motivated for both groups and during both sessions. 

\subsubsection{Tasks load}
As seen in the second sub-figure in Figure\ref{fig:res-ui}, the average tasks load scores for the two sessions tended to be slightly lower for the experimental group, suggesting that even with removing the agent’s support, the task of generating divergent questions was perceived to be easier for participants of this group.

\section{Discussion}
In this work we investigate the efficiency of a conversational agent designed to foster curiosity-driven learning in children. For our data analysis, we rely on behavioral measures of epistemic curiosity such as the ability to generate divergent questions and to lead autonomous explorations.

Our findings showed that our incentive agent was more successful in enhancing divergent question-asking abilities and transferring them to new contexts where no help/ cues are given to help participants. We could therefore argue that children of the experimental group were more able to maintain their performance level not only because the agent was successful in training the formulation of divergent questions but, even further, because it was successful in training their abilities to identify knowledge gaps while learning. Indeed, this skill is considered as the fuel for an individual's curiosity-driven behaviors: not being able to identify missing information is considered as the first brake that keeps children from asking questions in classrooms \cite{Graesser0}.

In addition, our findings showed a better exploratory performance for the experimental group, with a significant impact of the divergent question-asking performance on the exploration time. The incentive agent's success in promoting longer question-guided explorations could be traced back to its behavior that relied on giving time and encouraging children to think of missing information, and write them down, while exploring. Indeed, this is similar to using a 'Thinking aloud' approach that capitalizes the idea that verbalizing one’s thinking process explicitly while trying to solve a problem may lead to more carefully-planned problem-solving steps (\cite{Henjens2007}, \cite{Bannert2008}). 
Thereby, and together with the QA-training session, we could suggest that our intervention succeeded in giving participants the essential metacognitive tools they needed to pursue a cycle of curiosity-driven learning: a training to get used to stopping and taking time to think of a potential knowledge gap (the exploration session), and a training to identify a knowledge gap and express it (the QA-training session).

These results are also in accordance with literature in meta-cognition that has repeatedly shown that learners often tend to end their learning cycles prematurely because they overestimate their mastery of the skills (\cite{Kornell0}, \cite{Murayama1}). With our findings, we can suggest that training the faculty to stop, think of and ask divergent questions while learning might be a valuable strategy to be fostered in children in order for them to avoid the 'illusion of knowledge' trap and to prevent them from interrupting their information-seeking cycles prematurely.

As for the learning progress, we find that the experimental group was more successful and that our main behavioral curiosity measure (i.e. divergent question-asking performance) had a significant impact on it. This finding reinforces work such as in \cite{oudeyer1} that suggests a causal and bi-directional link between curiosity and learning. It is also in accordance with studies such as in \cite{Kang1}, where authors show that states of curiosity can enhance learning and memory retention. Indeed, we can see that, because our incentive agent was more successful in eliciting participants curiosity, it helped them generate the appropriate divergent-thinking questions while exploring which, subsequently, led them to acquire new information. Finally, and as differences in the participants' learning progress were shown to be correlated to the differences in their divergent question-asking behaviors, we can suggest  that monitoring these latter could be a valuable indicator for teachers to prognosis their students' learning progress. 

Overall, and to the best of our knowledge, our study is the first that investigates the different components of a curiosity-driven learning loop: it brings an existence evidence to the links between knowledge-gap experiences, knowledge-seeking behaviors and learning progress. It also reinforces the support to the "information gap" theory.

As a final important observation, our results failed to show higher intrinsic motivation scores for the experimental condition. However, we noted stable and high general motivation scores, irrespective of the condition. This is counter our hypothesis that relies on the learning progress model where an increased learning progress should yield increased intrinsic motivations. Indeed, as our incentive agent generated more learning progress, we expected higher intrinsic motivation for its group. Several explanations can be advanced. First, and as mentioned in \cite{Graesser0}, the learning by asking questions strategy, especially using digital tools is still rarely used in classroom settings, even today. It is therefore possible that the activity's novel and playful characters alone have made it very motivating and attractive for children, with both agents. Second, the use of a participatory design approach could have had inter-played with this finding also. Such a design method is well documented to increase the match between the proposed solution and the learners' needs \cite{Iversen2017}. Indeed, this approach led us to gather pedagogical content that is both appropriate in terms of learning objectives and accessible in terms of presentation format. Such features can enhance the solution's effectiveness and the learners' intrinsic motivation \cite{Chau0}. Finally, another possible explanation is children's incapacity to self-assess their learning progress, due to their metacognitive immaturity \cite{VanLoon2019}. In fact, learning judgment failures would reduce the reinforcing power of learning progress on intrinsic motivation. 

With rather encouraging results, our curiosity-based interactive system succeeded in promoting knowledge-gap awareness through the practice of higher-level question-asking; a skill that helped participants shape their self-sustained exploratory behaviors and gain new knowledge.

\section{Limitations and future directions}
A limitation to this study is the lack of interactivity in the dialogue between the agent and the child. One direction to a future study is to have a more sophisticated agent that will be able to give children feedback about the questions they submit. We also aim to implement an agent that is able to to generate divergent questions so it can engage in a turn-taking interaction with the child. 
Another possible limitation of this platform may be its very accessible design. The less-formal presentation of the resources (audios, videos) could have formed favourable conditions for our incentive agent to work. This could give us general directions to replicate our protocol with different material and a less-accessible design.

Finally, one of our aims is to raise children's awareness about the importance of the curiosity-driven behaviors on their learning. But with this experiment, and even though we were able to see an improvement in these behaviors and in learning progress, we do not know if the intervention actually helped participants see the link between the two. A future direction for this work will be to add a space in the platform where users can find pedagogical tutorials and activities that explain how to use their metacognitive skills and curiosity-driven behaviors to boost their learning.

\section{Conclusion}
In this work, we contribute to the promotion of curiosity-based approaches that foster individualized learning, while exploring the social advantages of using artificial conversational agents. We show that curiosity-driven behaviors can be practiced with the agent and that these behaviors are related to the learning progress children can achieve. Our work and results motivate the implementation of such approaches both in classroom settings and e-learning environments.

\section*{Acknowledgements}
This work has been funded by the educational technologies start-up EvidenceB
and the French National Association of Research and Technology (ANRT). The authors also thank the
teachers who participated in this study and the research team members who helped conduct the
experiments in classes.

\bibliographystyle{plain}
\bibliography{article.bib}
\end{document}